\documentstyle[prb,preprint,aps]{revtex}

\begin{document}

\title{Point contact spectroscopy and temperature dependence of 
resistivity of metallic Sodium Tungsten Bronzes \\- Role of optical phonons}
          
\author{N. Gayathri\footnote[1]{email: gayathri@physics.iisc.ernet.in}}
\address{Department of Physics,\\ Indian Institute of Science, Bangalore 560012, India}
\author{A. K. Raychaudhuri\footnote[2]{email:arup@csnpl.ren.nic.in}}
\address{National Physical Laboratory,\\ Dr. K.S.Krishnan Marg, New Delhi, India}

\maketitle

\begin{abstract}
In this paper we report the results of electrical resistivity (1.5$\leq$ T $\leq$ 300K) 
and point contact spectroscopy (PCS) measurements on single crystals of metallic 
sodium tungsten bronze with varying sodium content. We have shown that the electron-
phonon coupling function as measured through PCS can explain quantitatively the
large temperature dependence of resistivity $\rho$ seen in these materials over the entire
temperature range. The electron-phonon coupling function shows predominately  large
peaks for phonon frequency range of 30 meV $\leq \omega \leq$ 100 meV which match well with 
the calculated optical phonons for WO$_6$ octahedron. The integrated
electron-phonon coupling constant $\lambda$ from this data is $\approx$ 0.25-0.45, depending
on the Na content.
\end{abstract}

\newpage

\section{Introduction}
Electrical conduction in oxides, in particular, in metallic transition 
metal oxides is a topic of current interest \cite{ARUP,TSUDA}. Over the last decade three  
classes of phenomena (namely superconductivity, colossal magnetoresistance 
and insulator-metal transition) have kept the oxides in the center stage. For 
the high T$_c$ cuprates even the normal state resistivity with a linear 
temperature dependence is an unresolved issue. While the problem is most 
severe in the high T$_c$ cuprates the temperature dependence of the 
resistivity in the other transition metal oxides have a number of 
interesting features in store. To be specific in this paper we are concerned 
with normal (i.e, non-superconducting) metallic oxides which show the 
$\rho$(T) as in a metal (i.e, d$\rho$/dT $>$ 0) and would like to             
investigate whether the temperature dependence of $\rho$ can be understood 
\underline {quantitatively}. In this context two important observations are noteworthy. 
First, in these oxides the resistivity even in the metallic state is quite 
high and second, the magnitude of the temperature dependence of resistivity, 
$\Delta\rho$(T)(= $\rho$(T)-$\rho_{0}$, where $\rho_{0}$ is the 
residual resistivity) is also very high. In fact, incomparison to 
conventional metals and alloys, where the $\Delta\rho$ (often          
$\leq$100$\mu\Omega$cm) arises 
from electron-acoustic phonon scattering, the $\Delta\rho$ in the 
metallic oxides can be considered giant. If the source of the extra                
$\Delta\rho$ in oxides is scattering of electrons by phonons then 
it is clear that there  
must exist a source of large density of phonons in these oxides. Such a  
contribution has been found in certain oxides.
It was shown in oxides like Ba$_{1-x}$K$_x$BiO$_3$ (superconductor with
T$_c\approx$17K at $x\approx$0.47) and Nd$_{1.85}$Ce$_{0.15}$CuO$_4$
(superconductor with T$_c\approx$ 22K) that the scattering from 
optical phonons in these materials make a significant contribution to the 
temperature dependence of $\rho$ in the normal state \cite{TRALSHA}. 
These studies obtained the phonon data from the superconducting tunneling 
spectroscopy which gives the Eliashberg function $\alpha^2F(\omega$) and 
used it to calculate the resistivity. This investigation is a very good 
beginning in the quantitative understanding 
of the temperature dependence of the resistivity of metallic oxides.
This study, however, had one big drawback. Most of the data were on     
polycrystalline thin film or bulk pellets. Also the samples used for the 
tunneling data were not the same as those used for the resistivity  
studies. In order to improve upon this, we did a complete investigation 
with single crystalline Sodium Tungsten Bronze system (Na$_x$WO$_3$) and used the same 
sample for both the resistivity measurement and the phonon spectra measurement 
from which the Eliashberg function $\alpha^2F(\omega$) can be obtained for calculation 
of the resistivity. In this investigation we take a fresh look into the problem in 
order to see that the large temperature dependence of the  resistivity 
seen in the metallic oxides indeed arise from the electron-optical phonon          
scattering. By making the  investigation on a different and somewhat 
simple metallic oxide system we also establish the generality of the phenomena. 

Our experiment contains essentially two distinct parts:
(1) measurement of the resistivity $\rho$ over the temperature range
1.5 $<$ T $\leq$ 300K and 
(2) measurement of the phonon spectra (i.e, $\alpha^2F(\omega)$) using the 
technique of point contact spectroscopy done at 4.2K. 
We then use the experimental phonon spectra to calculate 
the temperature dependence of the resistivity.

The point contact spectroscopy (PCS) was first used by Yanson \cite{YANSON} to 
obtain the phonon spectrum (PS) of conventional metals and dilute alloys from the 
non-linear current voltage characteristics of the point contact junction.
This spectroscopy has the main merit that, it can give the 
function $\alpha_{pc}^2F(\omega$) which is simply related to the Eliashberg function 
$\alpha^2$F($\omega$) (electron-phonon interaction function) for any 
metal including those with very weak electron-phonon coupling. 
The relation between $\alpha_{pc}^2F(\omega)$ and 
$\alpha^2F(\omega$) is essentially a factor which depends on the 
geometry of the junction formed between the tip and the sample
\cite{GRIM,YANSON2} as discussed later on. This 
method enables one to derive the phonon spectrum of those metals which 
are not superconducting. We have thus used this technique to obtain 
the $\alpha_{pc}^2F(\omega$) and use this function to fit the resistivity of  
the metallic sodium tungsten bronzes. It must be pointed out that this is 
the first time the phonon spectra is being studied using PCS technique 
in this type of metallic oxides.

\section{The Sodium Tungsten Bronze system}
The sodium tungsten bronze system, (Na$_x$WO$_3$), investigated in this
work is a 3-dimensional cubic perovskite oxide. The corner sharing 
WO$_6$ octahedra forms a 3-dimensional network in WO$_3$ which is an 
insulator 
because the conduction band formed by the overlap of oxygen 2p and W 5d 
orbitals is empty. On addition of Na, it contributes one electron per Na 
atom and the conduction band gets filled. 
For $x >$ 0.2 the system makes a transition to a metallic state. However,the 
cubic structure is stabilized only for $x >$ 0.4. This material is very 
similar to ReO$_3$ which may be the most metallic of all oxides and in 
single crystalline form can be comparable to copper\cite{TSUDA}. In our system 
for $x$ = 0.9 the residual resistivity, $\rho_0$ = 3$\mu\Omega$cm which is more   
metallic than many conventional metallic alloys. Thus varying $x$ (and hence 
the electronic concentration) one can go from a good metallic state to an 
insulating state. The sodium tungsten bronze system (Na$_x$WO$_3$) 
has been well studied because of this composition driven metal-insulator 
transition seen in this system \cite{EDWARDS,LIGHTSEY1,DUBSON,AKR}.
This system has an additional feature that when W is substituted by Ta,
the empty 5d orbitals of Ta compensates for  the electrons being            
introduced by Na. This reduces the carrier density and also introduces a large 
disorder. One can thus see the effect 
of the disorder on the resistivity much like one does in a conventional 
alloy. On introduction of Tantalum in the tungsten site the chemical formula can 
be written as Na$_x$Ta$_y$W$_{1-y}$O$_3$. 

To summarize, the tungsten bronze system used by us has 
three distinct advantages : (1) the oxide is a simple oxide with a cubic 
structure, (2) the metallic state can be tuned to different electron 
densities by changing x and (3) the Ta substitution allows a way to 
introduce disorder into the system. The above three features thus gives us
a good system in which we can make a comprehensive investigation.

\section{Experimental Details}
The samples studied were single crystals of composition $x$=0.9, 0.75, 
0.63 and one sample with Ta substitution with composition $x$=0.6 and $y$=0.1 
($x-y$=0.5). The single crystals were grown by 
electrochemical growth from molten bath of sodium tungstate and tungstic   
oxide. Details of the sample preparation and characterization are given in 
the reference \cite{DUBSON,LIGHTSEY2}. We essentially followed the same technique. 

The effective carrier concentration n$_{eff}$ as a function of $x$ 
or ($x-y$) is given by the relation $n_{eff}$ = $(x-y)/a_0^3$, where $a_0$ (the cubic
cell dimension) follows the empirical relation $a_0$=0.0820$x$+3.7845 $\AA$ \cite{BROWN}. 
The values of $n_{eff}$ calculated 
using the above relation and that obtained by Hall measurements \cite{MUHLESTEIN} are
tabulated in table I. n$_{eff}$(exp) is larger in comparision to n$_{eff}$(cal) 
by few tens of percent. In particular this discrepancy is larger for low n$_{eff}$. 
The reason for this discrepancy (while it is not very large) can be the assumption of the
free electron formula for obtaining n$_{eff}$ from the Hall coefficient.

The resistivity of the samples were measured from 1.5K to 300K using 
four probe low frequency bridge technique \cite{SPRB}. The 
contacts to 
the samples were made by first depositing Au and leads were attached 
using silver paint or silver epoxy. Pre-cleaning of the surface 
before deposition of the Au is very crucial in this case.
The dimension of the samples were 
less than 2mm x 2mm x 0.1mm. Because of the finite size of the 
contacts on the sample, the error in the measurement of the absolute 
resistivity is about 10\%. The resistances of the samples measured varied 
between 1.0m$\Omega$-5.0$\Omega$. We could measure the resistance value with a 
precision of 5$\mu\Omega$. 

The point contact spectroscopy was done at 4.2K in a dipper type
\cite{SRI} cryostat using Au as the tip to make the point contact. The tip 
was formed by electrochemical etching of Au wire 
using a solution of HCl, HNO$_3$ and HF in the ratio 3:2:1. 
We used an ac bias of 5-10 V for etching the Au wire. The I-V characteristics 
of the Au-sample point contact junction were taken by a ac modulation method. This
directly gives the conductance (dI/dV) of the junction as a function of V.
The bias used were upto 125mV. The junction was in contact 
with the liquid Helium bath at 4.2K and
hence there is no possibility of temperature drifts when the I-V
characteristics were obtained. Typical junction resistances were adjusted to
the order of 5-20 Ohms using a differential screw. The surface of 
the samples were carefully
cleaned just before mounting the sample and were subjected to minimum
exposure to the atmosphere to avoid contamination of the surface. 

\section{Results }
\subsection{Resistivity}
Figure~1 shows the resistivity of the samples as a function of 
temperature over the entire temperature range. The temperature dependence    
of the resistivity is like a metal and d$\rho$/dT $>$ 0.
It can be seen that the resistivity varies by a large amount by changing the 
value of x. In particular the Ta substitution can lead to a big jump in 
$\rho$. The value of the residual resistivity are given in table I. 
The residual resistivies of the samples agree well 
with the pervious reported values \cite{LIGHTSEY1,MUHLESTEIN}. 
In figure~2 we show the variation of $\rho$(4.2K)  
as a function of $x$ (for $x >$ 0.5)
obtained by other investigators \cite{ELLERBECK} on single crystals and data 
obtained from our single crystals. For samples with x$\leq$0.6 the residual 
resistivity generally varies within a factor of 2 depending on the extent of
disorder. The exact value of $\rho$(4.2)K, however does not affect our experiement
or the conclusions.

The temperature dependence till 300K (i.e, $\Delta\rho$(300K)) for the low 
resistivity samples are nearly 15-25 $\mu\Omega$ cm. While for the Ta 
substituted sample this can be as large as 88 $\mu\Omega$ cm. There is an 
interesting anomaly to the trend at $x$=0.75 so that the room temperature 
resistivity of this sample is actually lower than that of the $x$=0.9 
sample. This arises because of partial ordering of Na atoms as has been seen 
previously \cite{MUHLESTEIN}. We will see that this also affects $\Delta\rho$(300K).
A detailed discussion on this issue 
is beyond the scope of the paper. Using a free electron model we estimate the mean 
free path at the two temperatures. The elastic mean free path $l_0$ as well 
as the in-elastic mean free path $l_{in}$ at 
300K (giving rise to $\Delta\rho (300K)$) are shown in the table I. The elastic
mean free path varies strongly as $x$ is varied and for the Ta-substituted sample 
$l_0/a_0\approx $2. For the most metallic sample ($x$=0.9), $l_0/a_0 \approx$ 145. 
Thus in the range of the composition studied, the electrons are not on the verge of
localisation. Most importantly we are in a position to assess the strength of
inelastic scattering quantified by $l_{in}^{-1}$ in the presence of varied $l_0$
while keeping the system chemically unaltered. In comparision to variation
of $l_0$ with x the variation of $l_{in}$ is not that severe. Interestingly 
the Na ordering in $x$=0.75 leads to a drastic increase in $l_{in}$.
 
\subsection{Point Contact Spectra}
In figure~3 as an example we show the second derivative d$^2$I/dV$^2$ obtained 
from an experimental dI/dV-V curve for $x$=0.75.
We also show the dI/dV curve for the 
sake of completeness. Here dI/dV decreases as |V| is increased and it is a sign of
"metallic" point contact.
The voltage axis directly gives the phonon energy in eV 
and as stated earlier d$^2$I/dV$^2$ gives a measure of the phonon density
of states. Essentially d$^2$I/dV$^2$ arises from inelastic scattering of
electrons by the phonons. For the other samples the curves were similar and we 
donot show them to avoid over crowding of data.
As a comparison we show that data taken on Au (figure~3(a) inset) which acts as 
reference to show the differences in the PS of the two classes of metals. 
For conventional metals the maximum phonon energy is often $<$ 20 meV. 
For oxides there is a large density of phonons at energies beyond 20 meV.    
It will be   
our attempt to see whether the phonons seen in the point contact 
spectroscopy can give rise to the necessary electron-phonon scattering so 
that the temperature dependence of the resistivity can be explained. This 
requires a qualitative analysis of the experimental PCS data to obtain the spectral 
function  $\alpha_{pc}^2F(\omega$). It is important to note that the thermal 
smearing ($\approx$3k$_{B}T$=1 meV at 4.2K) is much small compared to the phonon 
energies which are relevant for our work. As a result we will not apply any 
correction to our experimentally obtained spectra and treat them as essentially
the zero temperature spectra.

\section{Analysis and Discussion}
\subsection{Analysis of Point Contact Spectroscopy data}
The phonon spectrum can be obtained from the second derivative of the 
I-V characteristics of the point contacts from the relation \cite{DUIF}

\begin{equation}
\frac{d^2I}{dV^2} = -\frac{2\pi e^3}{\hbar}\Omega_{eff}N(0){\cal F}(eV)
\end{equation}

\noindent where ${\cal F}(eV)$ is the spectral function related to 
the strength of the inelastic interaction which gives rise to the
features in d$^2$I/dV$^2$, $\Omega_{eff}$ is an effective 
volume of phonon generation within which the inelastic scattering takes place, 
$N(0)$ is the density of states at the fermilevel. When phonons cause the
inelastic scattering 
${\cal F}(eV)$ = $\alpha_{pc}^2F(\omega)$ which is related to the Eliashberg function 
$\alpha^2F(\omega)$
by a form factor \cite{YANSON2}. In the measurement of this electron-phonon
interaction by means of point contact, there is a considerable background which
is due to the scattering due to non-equilibrium phonons. Taking this into account
the final expression used for obtaining the $\alpha_{pc}^2F(\omega)$ is given by

\begin{equation}
\frac{d^2I}{dV^2} = const\times\left(\alpha_{pc}^2F(eV) + \kappa 
\intop_{0}^{eV}{\frac{\alpha_{pc}^2F(\epsilon)}{\epsilon} 
d\epsilon}\right) 
\label {pc}
\end{equation}

\noindent In the case of a heterocontact the spectrum is given by the sum of phonon 
spectrum (PS) in both the electrodes \cite{HETERO}. We have used Au as one of the 
electrodes and the PS of Au has features between 10 and 17 meV (see inset of 
figure~3(a)). By subtracting the PS of Au one can easily identify the PS of 
the other electrode. The value of the constant ($const$) depends on the regime 
(ballistic or diffusive) we are working in and this is characterized by the
Knudsen number $K$ $\equiv$ $l_0/r$, where $r$ is the diameter of the contact
region \cite{DUIF}. For hetero-junctions like ours where electrons in one electrode
(like Au) have a very large mean free path, the $l_0$ of the electrons in the electrode 
with higher $\rho_0$ limits the Knudsen number $K$. In table II the relevant junction
parameters are given. Thus the most metallic samples with $K>$1 ($x$=0.9 
and 0.75) lie in
the ballistic regime, the $x$=0.63 sample with $K$ $\sim$ 1 is in the transition regime
and the Ta substituted sample with $K$ $\ll$ 1 lies in the diffusive regime. 
The difference between the two cases comes in the effective volume of the contact 
$\Omega_{eff}$ in 
which the inelastic scattering takes place and which contributes to the non-linearity 
in the I-V characteristics \cite{DUIF}.
For $x$ = 0.63 and $x$=0.6;$y$=0.1 samples, the necessary corrections were applied to take into
account that the point contacts were diffusive. In the case of 
the ballistic regime, the volume $\Omega_{eff}$ = 8$r^3$/3 and in the
diffusive case $\Omega_{eff}$ = $\pi r^2 l_0$/4, where $l_0$ is the elastic mean
free path of the electron. All the relevant numbers of the point contact junctions
are given in Table II.
Eqn.~\ref{pc} was used to obtain the $\alpha_{pc}^2F(\omega$). The constant 
$\kappa$ was determined using the fact that the observed spectrum 
coincides with the background function for eV $> \hbar \omega_{max}$ 
\cite{HETERO}. $\omega_{max}$ was chosen to be 125 meV. The above integral equation
has been inverted to obtain the $\alpha_{pc}^2F(\omega)$ which will be the
total of the contributions from both the electrodes. Since $\alpha_{pc}^2F(\omega)$ 
is independent of the contact details, we have subtracted the PS of the Au from 
this data which contributes a small peak at $\omega \approx$ 10meV 
and 17meV as shown in the inset of figure~3(a). The resultant is the phonon spectra 
of the sample alone. In figure~4(a) 
we have plotted the phonon spectra obtained by inverting the d$^2$I/dV$^2$ 
data for $x$=0.9, 0.75, 
and 0.63. The curves have been shifted for the sake of clarity. It can be seen that 
there are distinct features at energies between 40-80 meV. We were able to resolve
the spectrum into 6 major peaks by using peak analysis tools. For one spectrum
the contributions are marked by dotted lines and peak positions by arrows.
The frequencies where the peaks occur were essentially
unchanged when we vary $x$, but there was a small but distinct increase in the spectral weights
of the peaks. 
In figure~4(b) we have shown the spectral weights ($\alpha_{pc}^2F(\omega)$) of 
some of the peaks marked in figure~4(a) for
$x$=0.9, 0.75 and 0.63. The peaks 4 and 6 show a constant increase as a function of
$x$ but peaks 2 and 5 show a dip at $x$=0.75. This sudden decrease at $x$=0.75 leads to
lower electron-phonon coupling constant and thus 
can be related to the low $\Delta\rho$ value observed for this composition 
(see table I). 

Before making any mathematical analysis we would first like to establish that 
the temperature dependence of resistivity $\Delta\rho$(T) or the inelastic 
mean free path $l_{in}$ really arises from the phonons particularly from those
in the spectra beyond 20meV. For most conventional metals the strongest contribution
of the phonon spectrum comes from phonons within $\omega \leq$ 20meV. 
The oxides thus differ from the conventional metals in the fact that 
a large density of vibrational states extend upto 100meV. The electron
phonon coupling constant $\lambda$ is given by \cite{GRIM}

\begin{equation}
\lambda = 2 \intop_{0}^{\omega_{max}}\frac{\alpha^2F(\omega)}{\omega} d\omega
\end{equation}

\noindent From the experimental curve we obtain $\alpha_{pc}^2F(\omega)$. 
We can connect $\alpha^2F(\omega)$ and $\alpha_{pc}^2F(\omega)$ by a factor $G$ 
($\alpha_{pc}^2F(\omega)$ = $G \alpha^2F(\omega)$).   
The factor $G$ depends on the contact characteristics and it can be
calculated knowing the mean free path $l_0$ and the contact dimension $r$ \cite{YANSON2}. 
For the ballastic regime the value is a constant $G$ = 0.25. For
the diffusive regime the value is given by the relation 
$G$ = (($K$-1)/2$K$)+(1/4$K^2$)$\times$ln(1+2$K$). The values obtained for our
samples are given in Table II. Using this corrected $\alpha^2F(\omega)$, we can obtain
the value of $\lambda$. If the inelastic
scattering of electrons arise from the phonons then $l_{in}^{-1}$ (which is a
measure of the scattering strength) should increase monotonically with $\lambda$.
In figure~5 we plot $l_{in}^{-1}$(300K) as a function of $\lambda$. This clearly
brings out that $l_{in}^{-1}$ is indeed closely related to $\lambda$. This 
confirms that the source of the inelastic scattering are the phonons in
particular those with $\omega >$ 20meV which have a very large spectral weight
in the range $\omega$ $\approx$ 40-80 meV. The value of $\lambda$ for weak 
coupling metals like Cu, Na etc. is $\approx$ 0.15. For strong coupling Pb,
$\lambda$ = 1.34. The oxide superconductors mentioned in section I have a $\lambda$ of
about 1.0. Comparing the values of $\lambda$ obtained in our samples, we can say that
the electron-phonon coupling strength is weak to intermediate ($\approx$ 0.2-0.4).
In the next section we make a 
quantitative evaluation of $\rho$ from the phonon data and compare it with the
experiment.

\subsection{Analysis of resistivity using the PCS}
The temperature dependence of the resistivity $\Delta\rho$(T) can be 
calculated using
the formula derived by Ziman, in terms of the 
$\alpha^2_{tr}$F($\omega$) (electron-phonon transport coupling   
function) \cite{GRIM}

\begin{equation}
\Delta\rho(T) =  \frac{16\pi^2}{\omega_p^2 k_BT} \times                       
\intop_{0}^{\omega_{max}}{\frac{\hbar \omega \alpha^2_{tr} 
F(\omega)}{(exp[\hbar \omega/k_BT]-1)(1-exp[-\hbar \omega/k_BT])}}
d\omega \label{zires}
\end{equation}

\noindent where $\omega_p$ is the plasma frequency. The transport coupling function 
differs from the 
Eliashberg function $\alpha^2 F(\omega)$ only by a factor analogous 
to the relation between $\alpha_{pc}^2 F(\omega)$ and $\alpha^2 F(\omega)$. Work 
by Allen and co-workers \cite{ALLEN} provides a basis for using $\alpha^2 
F(\omega)$ instead of $\alpha^2_{tr} F(\omega)$ in the above formula. We thus
replace $\alpha^2_{tr}F(\omega)$ by $\alpha^2 F(\omega)$ in eqn.\ref{zires}. 
As mentioned in the previous section, 
$\alpha^2 F(\omega)$ = (1/$G$)$\alpha^2_{pc} F(\omega)$, hence 
using the value of $G$ given in table II, we can 
evaluate the temperature dependence of the resistivity 
using eqn~\ref{zires}. The \underline {only} unknown parameter in eqn~\ref{zires} is the 
plasma frequency $\omega_p$ which we use as a fit parameter.

The resistivity obtained for the samples, were fitted using the above 
relation for the temperature dependent part. The fit 
are given in figure~6(a)-(d). The maximum fit error are given in table III. 
It is interesting to see that the data can be
fitted with the experimental $\alpha_{pc}^2F(\omega)$ over the whole temperature
range for all the samples. The maximum fit error $\leq \pm $0.5\%. The fit error was
random and showed no systematic deviation either at low T or high T. The fit thus 
can be considered excellent because the only free 
parameter available is the plasma frequency $\omega_p$ which appears as a 
multiplicative factor only. The prefactor $\omega_p$ obtained in the above
relation from the $\rho$ data is listed in table III ($\omega_p$(exp)). 
The table also contains 
the values of $\omega_p$ as observed experimentally by electron energy loss spectroscopy
(LEELS) which we call $\omega_p$(obs) \cite{HILL}.

For all the samples the value of $\omega_p$ obtained from the
resistivity data $\omega_p$(exp) (eqn~\ref{zires}) and that obtained from LEELS 
$\omega_p$(obs) agree to better than
20\% -30\%. We think that this is a very good agreement given the fact that they
are obtained from two widely different techniques. There is, however, a systematic
difference. $\omega_p$(exp) is always lower than $\omega_p$(obs). One reason for this
could be that while $\omega_p$(exp) is obtained from the bulk measurements,
$\omega_p$(obs) is obtained from the surface sensitive LEELS. In these materials the Na has a
tendency to diffuse to the surface. This makes the surface somewhat Na rich compared
to the bulk and thus the n$_{eff}$ at the surface can be somewhat
larger than that in the bulk. We feel this makes $\omega_p$(obs) systematically larger
than $\omega_p$(exp).  

For the sake of completeness we would like to investigate whether instead of analysing 
$\rho$(T) using the complete phonon spectra we use a discrete frequency approach. In
this case we take a Bloch-Gruneisun type of electron-acoustic phonon interaction
characterized by a Debye temperature $\theta_D$ \cite{ZIMAN} and an electron-optic phonon 
contribution characterized by a temperature $\theta_E$ \cite{OPT}. The resulting $\rho$ is given
by

\begin{equation}
\rho(T) = \rho_0 + A~\left(\frac{T}{\theta_D}\right)^5 
\intop_{0}^{\theta_D/T}{\frac {x^5}
{(e^x - 1)(1 - e^{-x})}} + B\frac{\theta_E}{T} 
{sinh^2(\theta_E/2T)}^{-1}  \label{res}
\end{equation} 

\noindent Results of such a fit is given in the table IV. A typical fit shown 
in figure~7 for
x=0.9 sample. For comparison we have plotted the fit error obtained using eqns 
\ref{zires} and \ref{res} in figure~8. 
It is clear that eqn~\ref{res} can be used to fit the $\rho$ vs T curve but the extent of
agreement is poorer compared to that obtained from eqn~\ref{zires}. The value of 
$\theta_D$ obtained is close to what one expects for these oxides. 
From the specific heat
data\cite{ZUMSTEG} on these materials the values of $\theta_D$ are $\approx$ 300 - 400K. 
The value of $k_B\theta_E \approx$ 55 - 80 meV. This corresponds very well to 
some of the peaks in the phonon spectrum (figure~4a). 

Though the values of $\theta_D$ and $\theta_E$ are reasonable, the resistivity
can definitely be better explained by the complete phonon spectrum rather than assuming
the above expression \ref{res} to be valid. 

\subsection{Effect of Ta substitution}
It can be seen from figure~1 that even a small (10\%) substitution of W by Ta
leads to a large change in $\rho$(T). Between the $x$=0.63 and $x$=0.6 $y$=0.1 samples, 
n$_{eff}$ differ by $\approx$ 20\%. But the change in $\rho_0$ is by a factor of 
more than 6 and that in $\Delta\rho$(300K) by a factor of nearly 2. The empty Ta
orbitals at random sites thus act as a strong source of disorder in the system
leading to substantial electron scattering which brings down the elastic mean 
free path $l_0$ to just twice the cubic unit cell dimension. The comparision
of $x$=0.63 and $x$=0.6 $y$=0.1 sample will thus be like comparing a "metal" and a
"substitutional alloy". 

Our concern in this paper is the temperature dependent part $\Delta\rho$(300K) which changes 
by nearly a factor of 2 and the inelastic mean free path $l_{in}$(300K) which is
also quite small ($\approx$ 27$\AA$) in the Ta substituted samples. 
We would like to see whether the
changes to $\alpha^2F(\omega)$ brought forward by Ta substitution can also explain
the temperature dependent part. This is an important test of our work that the 
$\Delta\rho$(300K) can be quantitatively explained by the observed $\alpha^2F(\omega)$
even when there is large substitutional disorder and the elastic mean free path is
very small.

In figure~8 we show the $\alpha_{ph}^2F(\omega)$ for the two samples. 
Ta has very similar mass as W. In the region $\omega \approx$40 - 60 meV there is a
large enhancement of $\alpha^2F(\omega)$ and the peaks have moved towards 
higher $\omega$.
The enhancement of the $\alpha_{ph}^2F(\omega)$ in
the region $\omega \approx$ 40 - 60 meV makes the $\lambda$ go up by a factor of 1.5. 
This causes the $l_{in}$ to go down by a factor of 1.75. Thus we see that the
resistivity enhancement $\Delta\rho$(T) can be explained by the enhancement of 
electron-phonon interaction in the Ta substituted sample. This can be put into a simple
description that whenever a large change occurs in the phonon spectrum a comparable
change is also seen in the inelastic mean free path. Our above extensive analysis
quantitatively establishes the fact that the temperature dependence of $\rho$ in these
oxides can be explained by the electron-phonon interaction. These modes mostly
occur in the range 40 - 100 meV. We next discuss the origin of these modes and establish 
that they are optic modes.

\subsection{The optic modes}
WO$_3$ is a polar crystal which has a basis consisting of differently charged ions. When
Na is added, the empty conduction band is filled but the polar basis lattice remains
unchanged.
Na$_x$WO$_3$ can be considered as a polar metal, which we can 
define as polar substance (WO$_3$) with a partially filled conduction 
band. The structure of the WO$_3$ (ReO$_3$ structure), is a vacant 
structure, where the oxygen ions can oscillate with large amplitudes 
(lower energy). The frequency of this 
oscillations correspond to the optical modes of the lattice 
vibrations. In a polar crystal, 
the optical vibrations are accompanied by a displacement 
and a deformation potential leading to polarisation. The total 
polarisation leads to a displacement which causes a 
strong coupling between the electrons and the optical phonons 
\cite{JPC83}. The optical mode in WO$_3$ corresponds to a temperature of 
600K or 50 meV \cite{JCP63}. As Na is doped into the 
system, the Na occupies the vacant space and the amplitude of the 
oscillation of the oxygen ions decreases. This increases the energy 
corresponding to the oscillations. This shifts the spectral weight of the
different peaks towards higher values of $\omega$ (see figure~4a). The phonon
frequencies of  Na$_x$WO$_3$ mostly arise from the WO$_6$ octahedra and thus
are similar to WO$_3$. The phonon dispersions in WO$_3$ has been calculated \cite{JPC83}
and we use this as an identification guide to our experimental spectra. In WO$_3$ the
highest calculated acoustic mode frequencies (at the zone boundary) is $\approx$ 25 meV.
The lowest peak in $\alpha^2F(\omega)$ occuring below 30 meV (i.e, peak \#1) is likely
to arise from the acoustic phonons. The lowest calculated optical mode frequency  (at zone
center) $\approx$ 33 meV. Thus all the peaks of $\alpha^2F(\omega)$ for $\omega >$ 30 meV
are from optical phonons. At the zone center the calculated phonon frequencies are 33  meV and
77 meV for $\omega <$ 100 meV. At the zone boundary, the phonon frequencies are calculated
to be around $\omega \approx$ 55, 75, 85, 115 meV where the exact position depends on the 
symmetry directions. The calculation of electron-phonon interaction parameters
show that it is wave vector ($q$) dependent and also symmetry dependent \cite{JPC83}. 
The resulting $\alpha^2F(\omega)$ will have to be determined by an integration of all $q$ using
the phonon dispersion curve. Such a calculation is beyond the scope of this paper.
We only note that in most symmetry directions the calculated electron-phonon
interaction parameters reaches a high value for $q\pi$/a$_0 >$ 0.5 - 0.75. As a result peaks
in $\alpha^2F(\omega)$ are expected to occur for values of phonon frequencies 
intermediate between those calculated for the zone center ($q$=0) and zone boundaries
($q\pi$/a$_0$=1). We find that the observed peaks in $\alpha^2F(\omega)$ indeed
occur between these limits. Also the splitting into longitudinal and transverse optical
modes leads to a maximum of 5 distinct frequencies in the range 30meV$<\omega<$100meV.
This also matches with our observation. This establishes clearly that the peaks in 
$\alpha^2F(\omega)$ indeed are related to the indentifiable optical phonon frequencies
of the WO$_3$ structure. In insulating WO$_3$ the electron-optical phonon coupling is
quite strong due to the polar nature of the material. However, on introduction of Na as 
$x$ increases the material becomes conducting which leads to screening of the interaction
which reduces the electron-optical phonon coupling.

The Ta substitution in place of W leads to an increase in the $\alpha^2F(\omega)$ for
$\omega$ lying in the range 40-60 meV. The $\alpha^2F(\omega)$ remains unaltered in the
other frequency range. This is rather interesting and allows us to draw some 
interesting conclusions. The phonon dispersion calculations show that the lowest
optical phonon branches describe a displacement field in which the W(or Ta) atom
moves in one direction and the oxygen atoms move in the other direction. This leads
to a large dipole moment and large electron-lattice coupling. This will therefore
depend strongly on the charge on the W-site and the screening associated with the
atomic charges. When Ta is substituted for W, there is a hole on the W-site. This
reduces significantly the screening and the coupling constant is strongly enhanced.
This explains the increase in the value of the $\lambda$ in the case of Ta substituted
sample ($x$=0.6 $y$=0.1) compared to the $x$=0.63 sample.

\section{Conclusion}In this paper we have brought out the role of
optical phonon modes of simple perovskite oxides in the resistivity.
The phonon spectra, better still the electron-phonon spectral frequencies were
directly measured by point contact spectroscopy.
We find that, there are distinct phonon modes at enegies between 40-80 meV in
the phonon spectrum ($\alpha^2 F(\omega)$) obtained using point contact spectroscopy.
Using this phonon spectrum we are able to explain the temperature dependence of 
resistivity of the metallic sodium tungsten bronzes. We have obtained the values
of $\lambda$ the electron-phonon coupling constant. Comparing this with the
values obtained for other conventional metals and superconductors, we can conclude 
that the electron-phonon interaction in this system is in the weak coupling limit. It 
may not induce superconductivity but this
has significant influence on the temperature dependence of resistivity. The paper
is quantitative and clearly shows that the large temperature dependence of $\rho$
seen in these class of oxides essentially arises from interaction with optical
phonons.

\newpage
{\center \bf FIGURE CAPTIONS}

\noindent{\bf Figure 1} Resistivity of the Na$_x$Ta$_y$W$_{1-y}$O$_3$ samples.

\noindent{\bf Figure 2} $\rho_{4.2K}$ values obtained from our investigation and Ellerbeck et.al. \cite{ELLERBECK}

\noindent{\bf Figure 3} Point contact spectra. {\bf(a)}d$^2$I/dV$^2$ for $x$=0.75 obtained at T=4.2K. 
{\bf Inset} shows the Phonon spectra of Au.
{\bf(b)} dI/dV obtained by modulation technique for the same sample (R$_0$ $\approx$ 5$\Omega$).

\noindent{\bf Figure 4} {\bf(a)} The point contact electron-phonon interaction function 
$\alpha_{pc}^2F(\omega)$ obtained using eqn~\ref{pc}. The peaks have
been marked following the multiple peaks analysis. {\bf(b)} The spectral intensity
of some of the peaks as a function of composition $x$.

\noindent{\bf Figure 5} The inverse of inelastic mean free path at 300K ($\l_{in}^{-1}$) as a
function of the electron phonon coupling strength ($\lambda$).

\noindent{\bf Figure 6(a-d)} The fit to the resistivity using eqn.\ref{zires} for the samples. The
only fit parameter is $\omega_p$ tabulated in table II.

\noindent{\bf Figure 7} {\bf(a)}Fit to the resistivity using eqn.\ref{res} for $x$=0.9 sample. {\bf(b)}
Fit error (\%) for the two fitting schemes ((a)
eqn.\ref{zires} and (b)eqn.\ref{res} for $x$=0.9 sample.

\noindent{\bf Figure 8} $\alpha^2F(\omega)$ for $x$=0.63 and $x$=0.6;$y$=0.1 samples. The difference
in the spectral intensities clearly indicate greater electron-phonon interaction 
strength (characterized by $\lambda$) for the Ta substituted sample.

\newpage
{\center TABLE I}\\
\begin{center}
\begin{tabular}{|c||c|c||c|c|c|c|}
\hline
Composition & n$_{eff}$ (calc) & $n_{eff}$ (exp) \cite{MUHLESTEIN} &
$\rho_0$ & $\Delta\rho$(300K) & $l_{0}$ & $l_{in}$(300K) \\
$x$ \hspace{0.5cm} y & /cm$^{3}$ & /cm$^{3}$ &
$\mu\Omega$ cm & $\mu \Omega$ cm & $\AA$ & $\AA$ \\
\hline
0.9 \hspace{0.5cm} -& 1.57$\times$10$^{22}$ & 1.95$\times$10$^{22}$ &
3.20 & 25.3 & 550 & 70 \\
0.75 \hspace{0.5cm} -& 1.32$\times$10$^{22}$ & 1.60$\times$10$^{22}$ &
6.95 & 15.3 & 288 & 130 \\
0.63 \hspace{0.5cm} -& 1.12$\times$10$^{22}$ & 1.49$\times$10$^{22}$ &
47.01 & 42.0 & 45 & 47 \\
0.6 \hspace{0.5cm} 0.1 & 8.91$\times$10$^{21}$ & 1.18$\times$10$^{22}$ &
330.10 & 88.0 & 7.5 & 27 \\
\hline
\end{tabular}
\end{center} 

\vspace{1.5cm}
{\center TABLE II}\\
\begin{center}
\begin{tabular}{|c|c|c|c|c|}
\hline
Composition & R$^{\small{\dag}}$ & $K^{\small{\ddag}}$ & $G^{\small{\ast}}$ 
& $r^{\small{\flat}}$ \\
$x$ \hspace{0.5cm} y & $\Omega$ &  & &  $\AA$ \\ \hline
0.9 \hspace{0.5cm} - & 5.0 & 5.0 & 0.25 & 110 \\
0.75 \hspace{0.5cm} - & 3.0 & 1.7 & 0.25 & 134 \\
0.63 \hspace{0.5cm} - & 15.0 & 0.5 & 0.19 & 90 \\
0.6 \hspace{0.5cm} 0.1 & 18 & 0.09 & 0.053 & 83 \\
\hline
\end{tabular}
\begin{quote}
$\dag$  : Junction resistance\\
$\ddag$ : Knudsen number \\
$\ast$  : Correction evaluated using expressions from ref[6]\\
$\flat$ : Contact dimension calculated using $r = (4\rho l/3\pi R)^{1/2}$ for 
          Ballastic regime and Wexler's interpolation formula for the diffusive
          regime. ($\rho$ and l are of the sample, since the contact resistance is
          limited by the sample because of it higher resistivity).
\end{quote}
\end{center}

\newpage
{\center TABLE III}\\
\begin{center}
\begin{tabular}{|c|c|c|c|}
\hline
Composition & $\omega_p$(exp) &  $\omega_p$(obs) \cite{HILL} & Fit error\\
 & eV & eV & \\
\hline
$x$=0.90      & 2.58 & 3.14 & $\pm$ 0.4\% \\
$x$=0.75      & 2.83 & 3.02 & $\pm$ 0.4\% \\
$x$=0.63      & 2.23 & 2.97 & $\pm$ 0.5\% \\
$x$=0.6 y=0.1 & 2.13 & 2.77 & $\pm$ 0.3\% \\ 
\hline
\end{tabular} 
\end{center}

\vspace{1.5cm}
{\center TABLE IV}\\
\begin{center}
\begin{tabular}{|c||c|c|c|c|c|c|}
\hline
Composition & $\rho_0$ & $A$ & $B$ & $\theta_D$ & $\theta_E$ 
& Fit error\\
 & $\mu\Omega$ cm & $\mu\Omega$ cm & $\mu\Omega$ cm & K & K & \\ 
\hline
$x$=0.90      & 3.17   & 74.55  & 16.94 & 356.4 & 916.6 & $\pm$ 4.0\% \\
$x$=0.75      & 7.00   & 42.37  & 7.09  & 334.6 & 756.1 & $\pm$ 2.5\% \\
$x$=0.63      & 47.05  & 67.21  & 18.62 & 252.4 & 645.7 & $\pm$ 2.5\% \\ 
$x$=0.6 $y$=0.1 & 330.14 & 296.19 & 33.94 & 328.2 & 855.2 & $\pm$ 1.5\% \\ 
\hline
\end{tabular}
\end{center} 

\newpage

{\center TABLE CAPTIONS}\\
{\bf Table I} The important parameters obtained from the transport measurements.

{\bf Table II} Parameters from the point contact junctions for all the samples used
in calculating the true $\alpha^2F(\omega)$.

{\bf Table III} Comparision of plasma frequencies obtained by our method and LEELS\cite{HILL} 
along with the maximum fit error obtained using eqn.\ref{zires} to fit the resistivity.

{\bf Table IV} Parameters obtained from the fit of resistivity to eqn~\ref{res}.

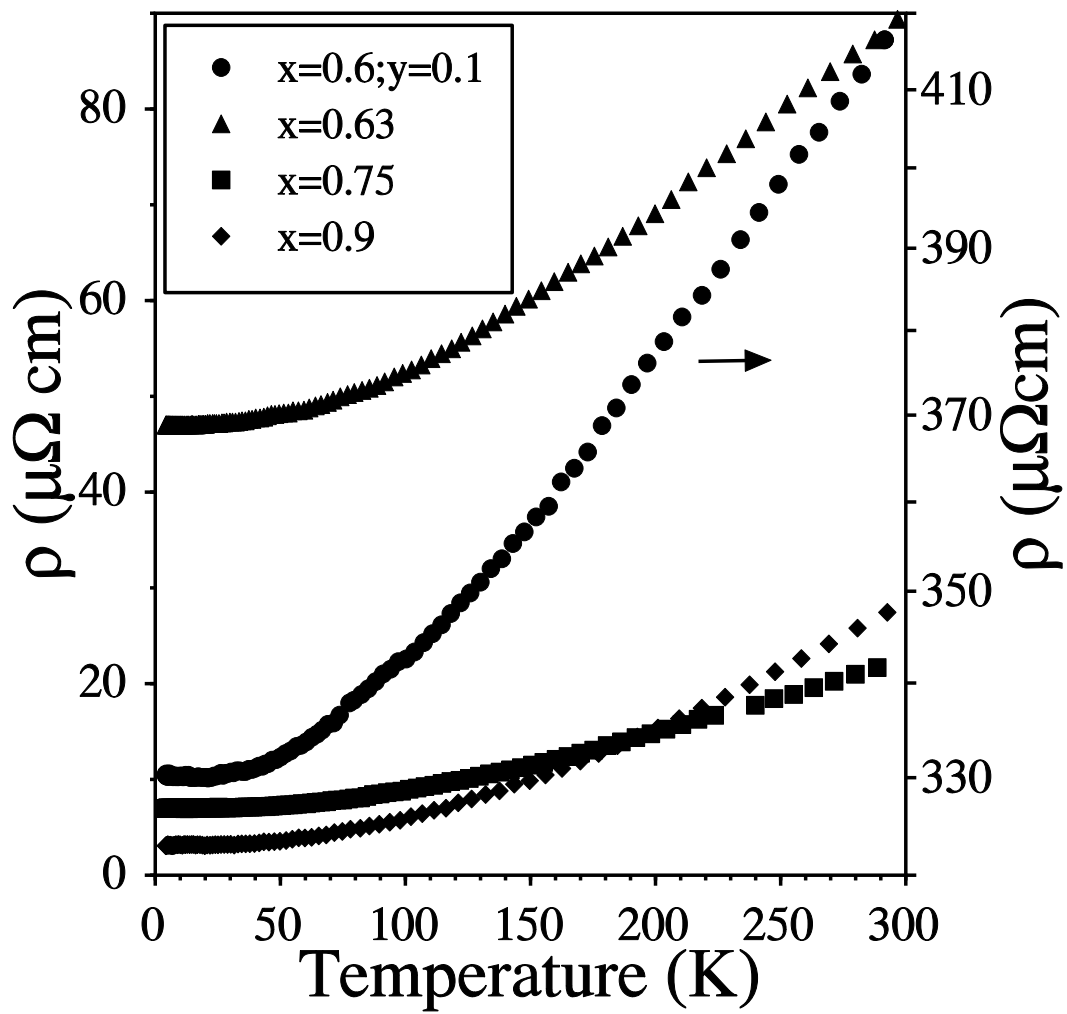
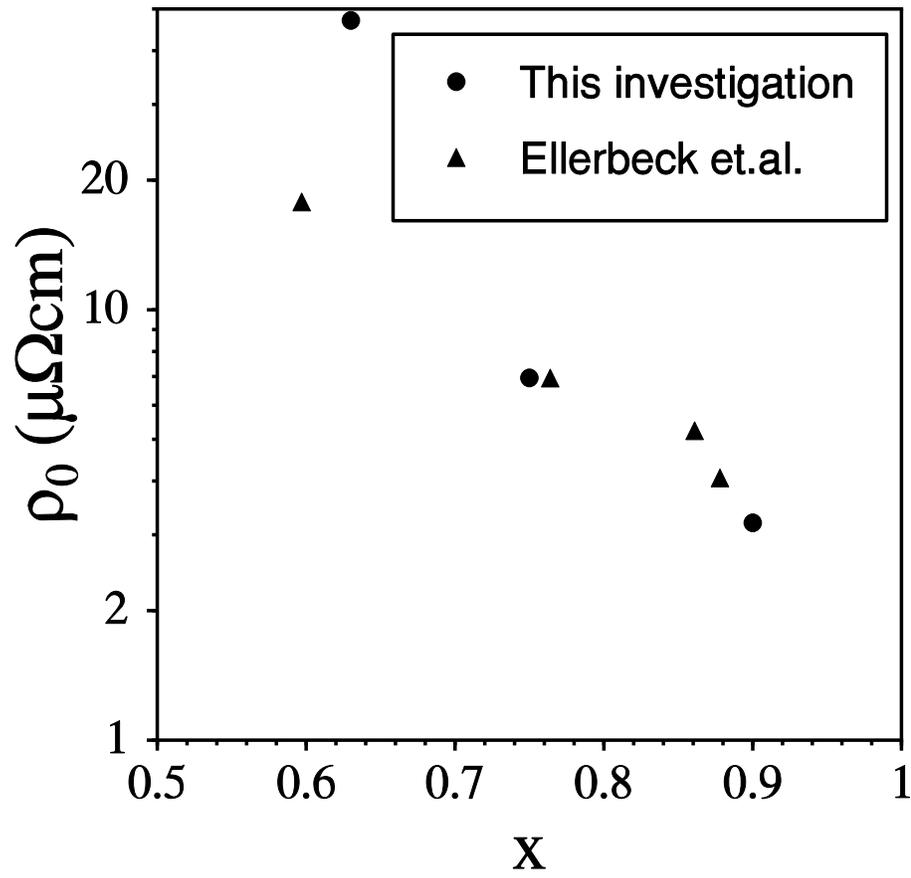
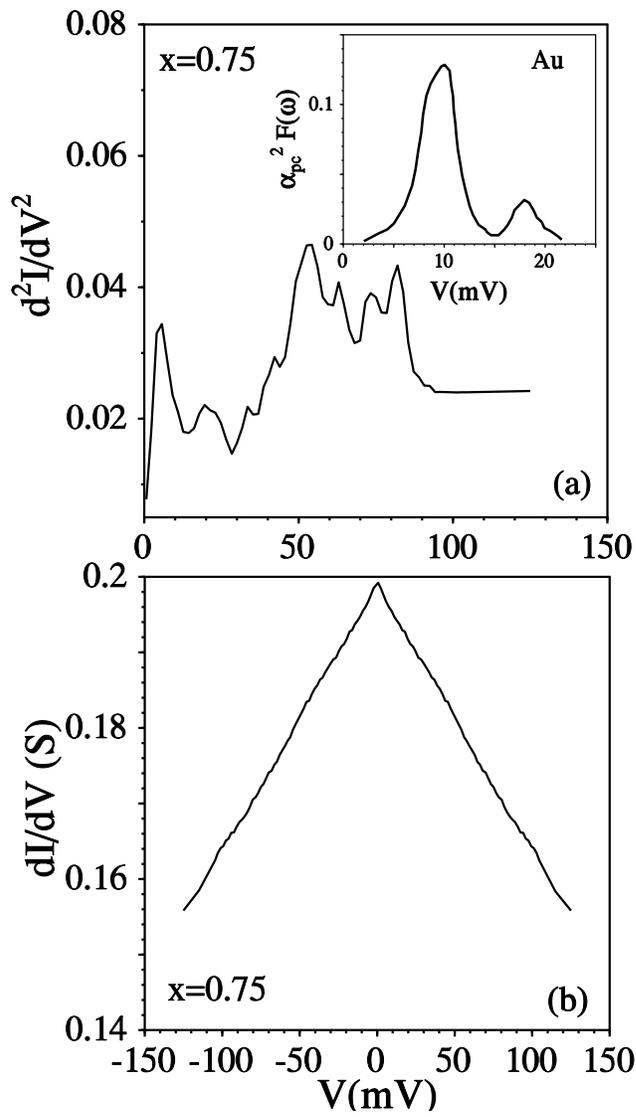
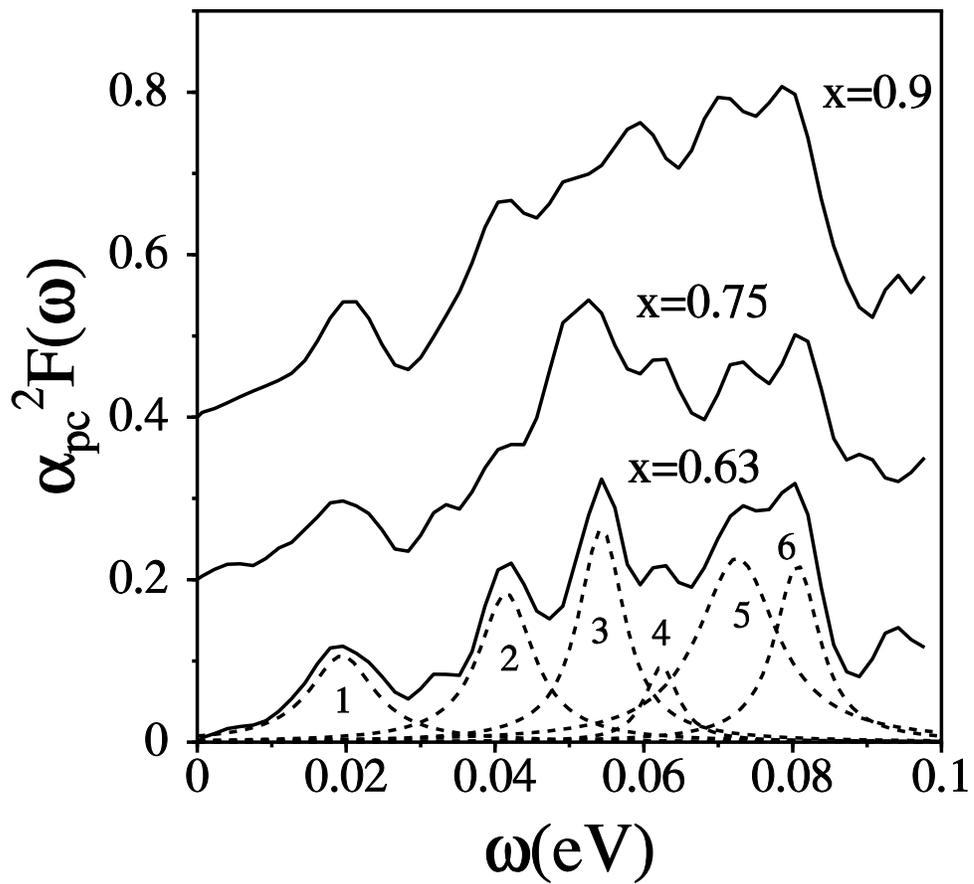
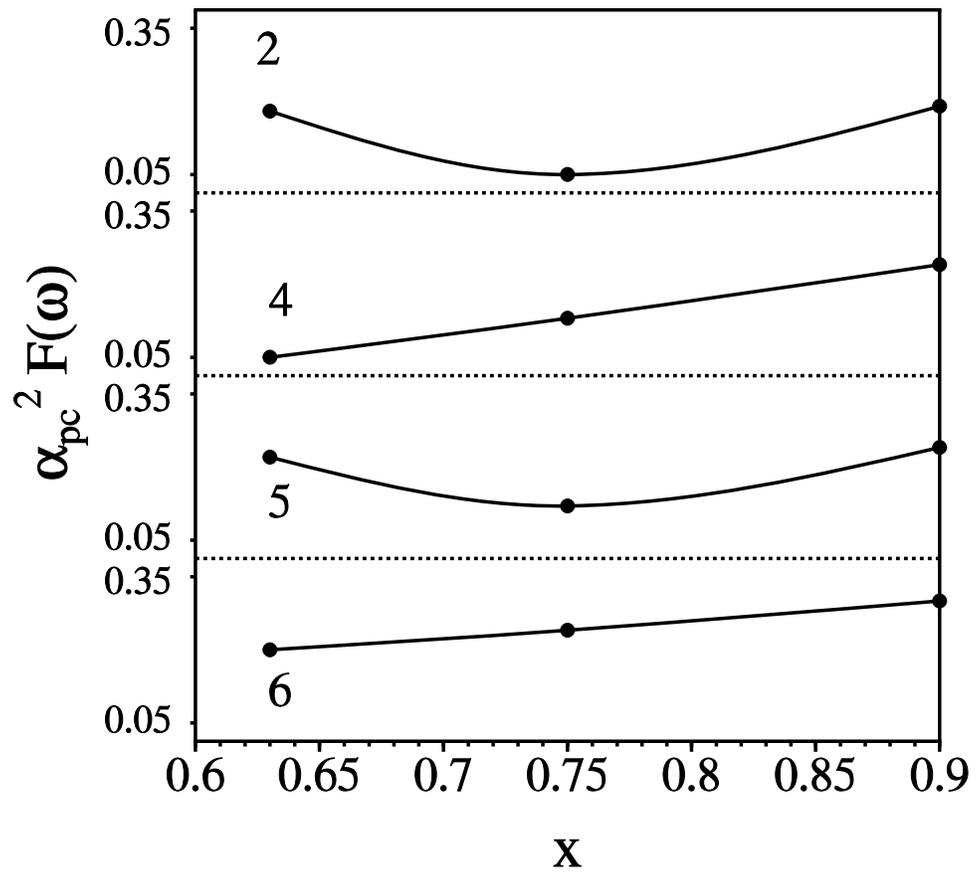
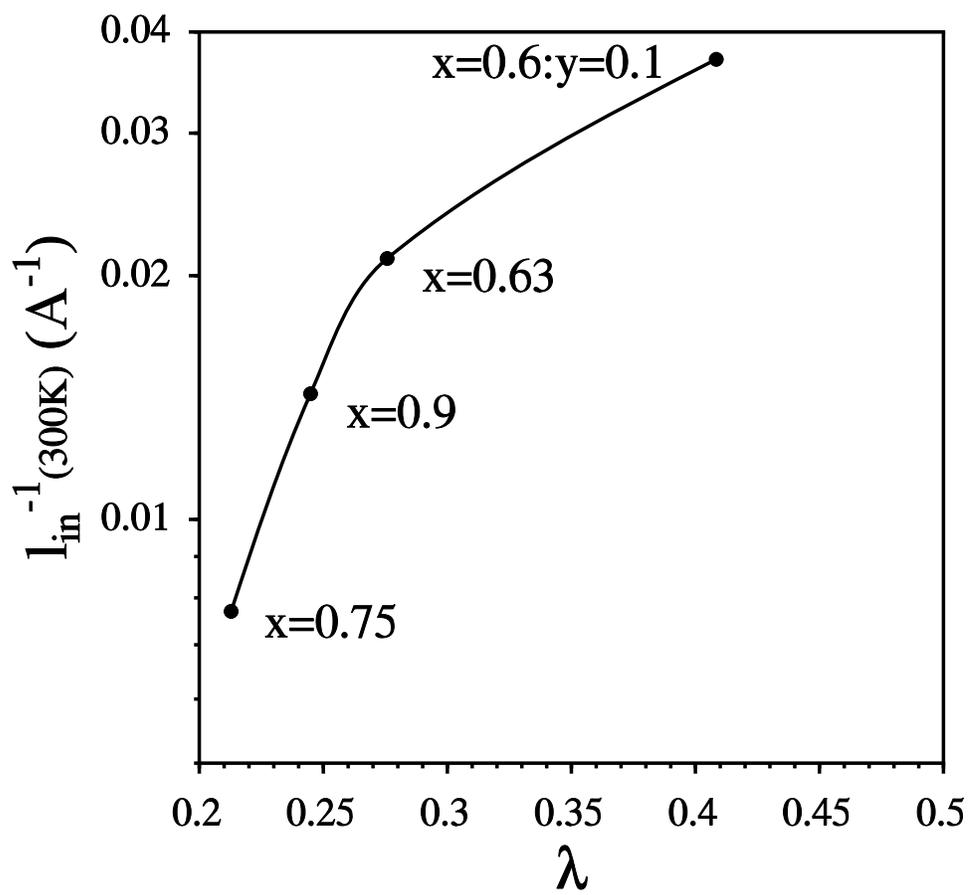
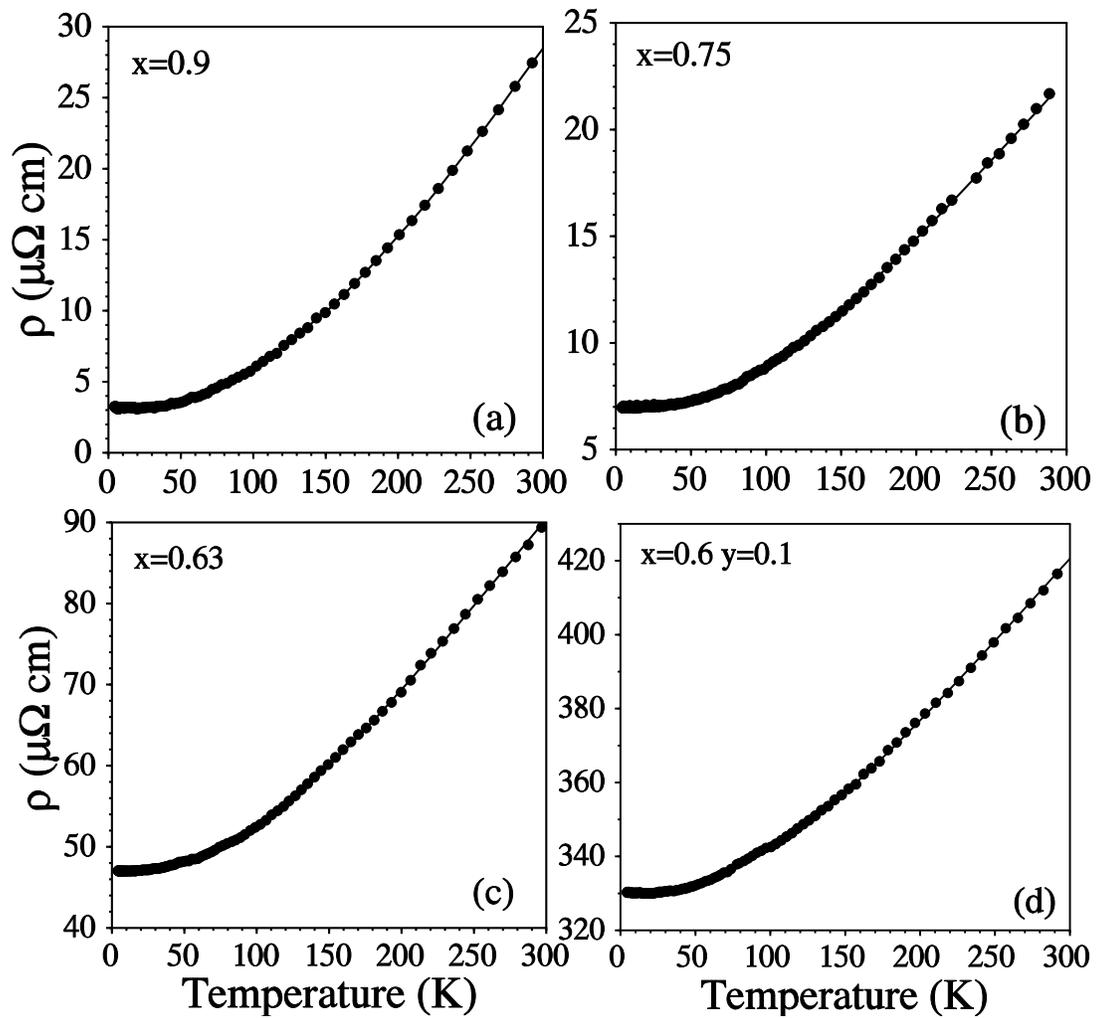
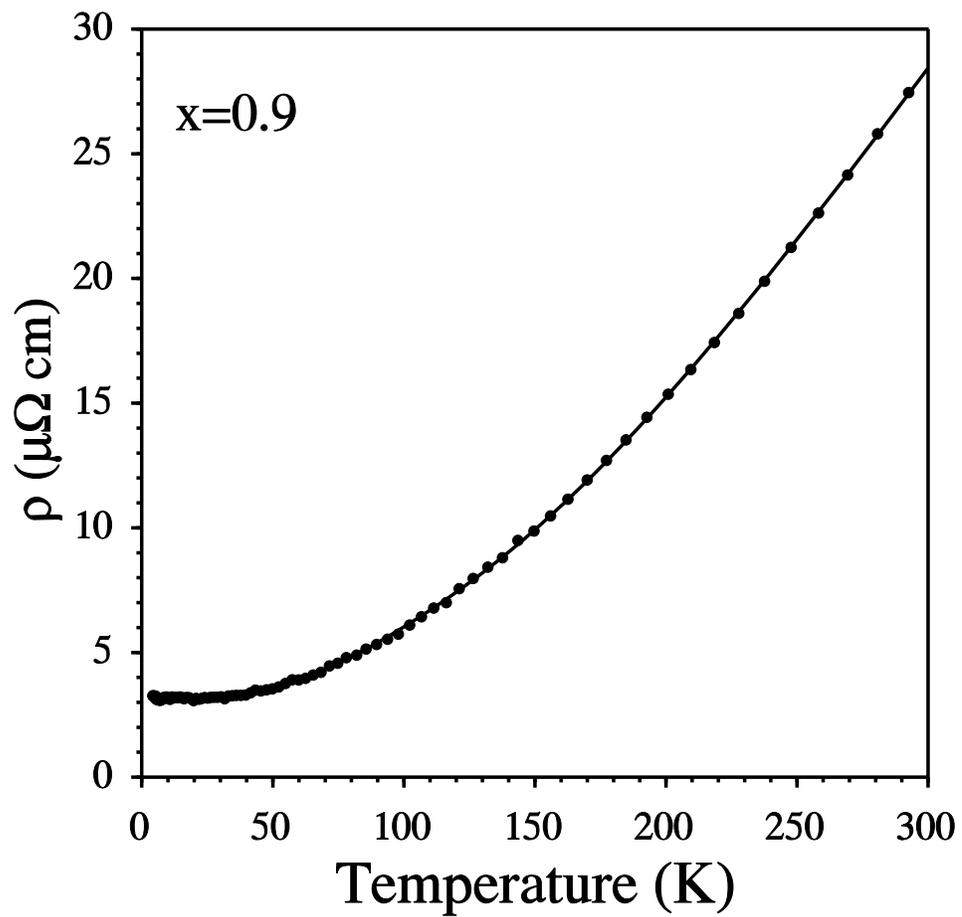
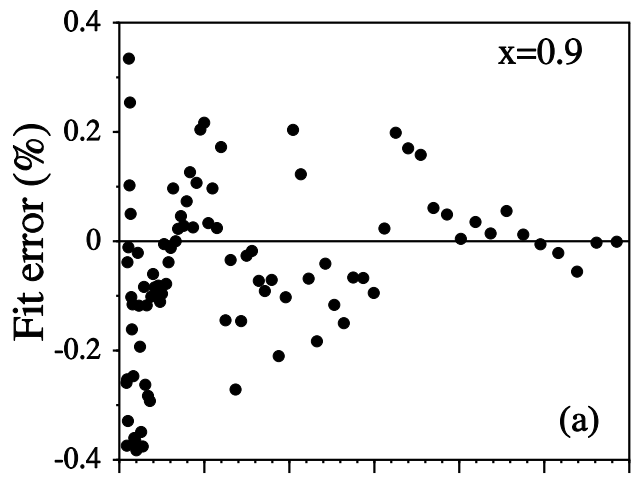
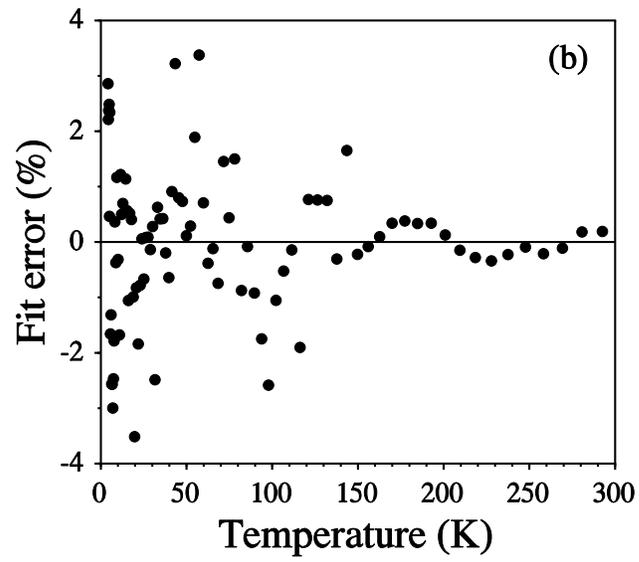
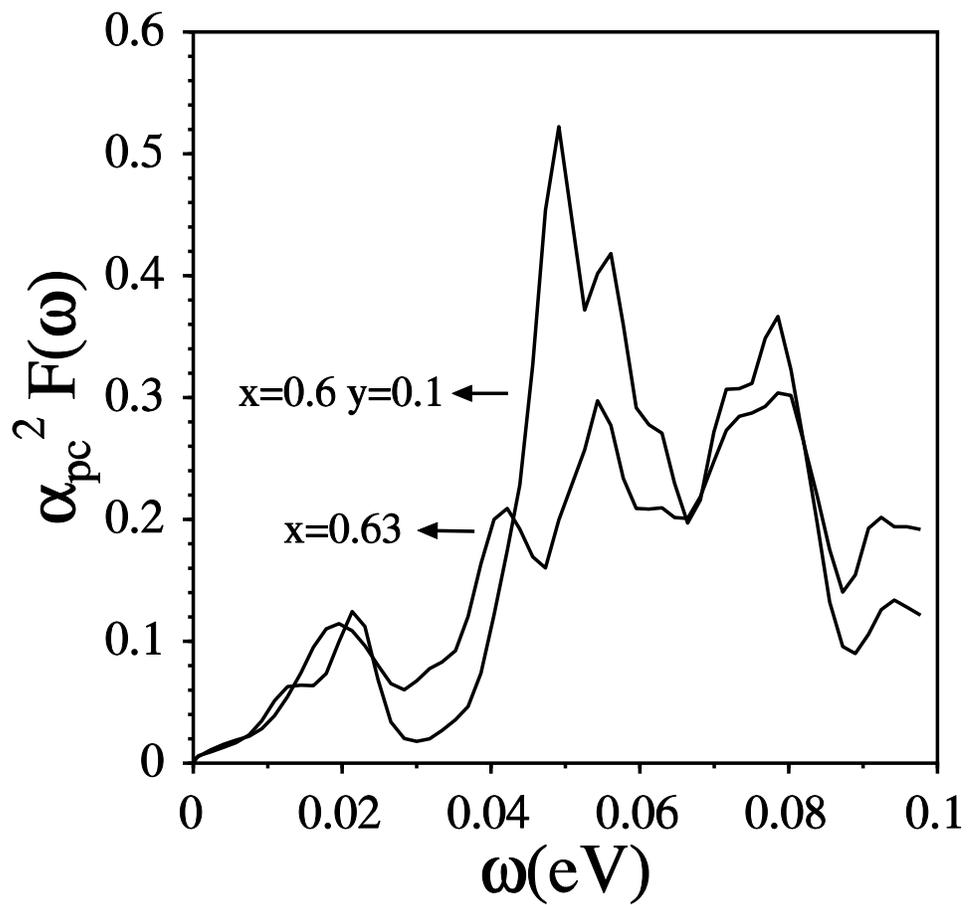
\end{document}